\documentclass[11pt,oneside,doublespace]{article}
\usepackage{geometry} 
\usepackage{amsmath,amsthm}
\usepackage{amssymb}
\usepackage{color}                   
\usepackage{rotating}
\usepackage{mdframed}
\usepackage{graphicx,amssymb}
\usepackage{algorithm}
\usepackage{algorithmic}
\usepackage{color}
\usepackage{diagbox}
\usepackage{multirow}
\usepackage{subcaption,appendix}
\usepackage{makecell}
\usepackage[affil-it]{authblk}
\usepackage{soul}

\usepackage{geometry}
\makeatletter

\@addtoreset{algorithm}{section}
\makeatother

\newtheorem{theorem}{Theorem}[section]

\newtheorem{definition}[theorem]{Definition}

\def\U{{\bf U}}
\def\V{{\bf V}}
\def\L{{\bf L}}

\def\G{{\cal G}}
\def\T{{\cal T}}

\def\R{ {\mathbb R} }

\numberwithin{equation}{section}

\setstcolor{red}
\title{Graph Fourier transforms on directed product graphs}

\author{Cheng Cheng, Yang Chen, Yeon Ju Lee and Qiyu Sun
\thanks{Cheng is with  School of Mathematics, Sun Yat-sen University,   Guangzhou, Guangdong 510275, China; Chen is with Key Laboratory of Computing and Stochastic Mathematics (Ministry of Education), School of Mathematics and Statistics,
 Hunan Normal University, Changsha, Hunan 410081, China; Lee is with Division of Applied Mathematical Sciences, Korea University Sejong Campus, Sejong City, 30019, Korea;
Sun is with  Department of Mathematics, University of Central Florida, Orlando, Florida 32816, USA.
Emails: chengch66@mail.sysu.edu.cn; ychenmath@hunnu.edu.cn; leeyeonju08@korea.ac.kr;  qiyu.sun@ucf.edu.
This work is partially supported by
the   National Nature Science Foundation of China (11901192, 12171490), Guangdong Province Nature Science Foundation (2022A1515011060),
the Ministry of Education of Korea grant funded by the Korea government (MEST) (No. 2021R1A2C1008360) and  National Science Foundation (DMS-1816313).
}}

\date{}

\begin{document}

	\maketitle

\begin{abstract}
 Graph Fourier transform (GFT) is one of the fundamental tools in graph signal processing to decompose graph signals into different frequency components and to represent graph signals with strong correlation by different modes of variation effectively. The  GFT on undirected graphs has been well studied and several approaches have been proposed to define GFTs on directed graphs. In this paper, based on the singular value decompositions of some graph Laplacians, we propose  two  GFTs  on the Cartesian product graph of two directed graphs. We show that the proposed GFTs  could represent spatial-temporal data sets on directed networks with strong  correlation efficiently, and in the undirected graph setting they are essentially the joint GFT  in the literature. In this paper, we also consider the  bandlimiting procedure in  frequency domains of the proposed GFTs, and demonstrate its performances  to denoise the  hourly temperature data sets  collected at 32 weather stations in the region of Brest (France) and at 218 locations in the United States.

\end{abstract}



	
	\section{Introduction}

Data sets in  many engineering applications are time-varying and
pairwise interactions among agents of a network are not always  mutual and equitable, such as the interaction data set on a social network
and the temperature data set collected by a weather observation network.
Those spatial-temporal data sets are usually modeled as graph signals residing on some directed product graphs
\cite{aliaksei14}-\cite{yamagata2022}.

 Graph Fourier transform (GFT)  
  is one of the fundamental tools to deal with spatial-temporal data sets
\cite{Singh16}-\cite{Yang2022}.
 The  GFT on undirected graphs has been well studied and a conventional definition is  based on
the eigen-decomposition of  the  
 Laplacian  
 on the graph
\cite{aliaksei14}, \cite{Chungbook}-\cite{stankovic2019introduction}.
 However, the above eigen-decomposition approach does not apply directly in the directed graph setting. In recent years,  several approaches have been proposed to define GFTs on directed graphs.



  The GFT  should be designed to decompose graph signals into different frequency components and to efficiently  represent them by different
modes of variation  \cite{Grassi18, ncjs22, Jiang22, Kurokawa17}.
The Jordan decomposition of the Laplacian has been widely used to define the GFT on directed graphs, but the computational cost
is high
and Parseval's identity may not hold 
\cite{aliaksei14,  deri2017, Domingos20, sandryhaila14}.
 Several directed
variations  of signals along the graph structure have been proposed to define GFT on directed graphs
 \cite{Singh16,Sardellitti17,deri2017, Marques2020}.
Based on the singular value decomposition (SVD)   of the Laplacian on directed graphs,
  the authors of this paper introduced a GFT on directed graphs  \cite{Yang2022}.
  The SVD-based GFT in \cite{Yang2022} has numerical stability and low computational cost, and on  directed circulant graphs
  it is
   consistent with the classical discrete Fourier transform. 

   Let ${\mathcal G}_1$ and ${\mathcal G}_2$ be two directed graphs of orders $N_1$ and $N_2$.
In this paper, we propose  two   GFTs ${\mathcal F}_\square$ and ${\mathcal F}_\otimes$ on the  Cartesian product graph ${\mathcal G}_1\square {\mathcal G}_2$ of two directed graphs
${\mathcal G}_1$ and ${\mathcal G}_2$, see  Definitions \ref{jgft1.def} and \ref{jgft2.def}. 
The proposed  GFTs 
are based on the SVDs of the Laplacians on the Cartesian product graph ${\mathcal G}_1\square {\mathcal G}_2$ and
 on directed graphs ${\mathcal G}_1$ and ${\mathcal G}_2$ respectively.
 We show that bandlimiting in the frequency domains of the proposed  GFTs ${\mathcal F}_\square$ and ${\mathcal F}_\otimes$
provide  good approximations to   signals on the Cartesian product  graph ${\mathcal G}_1\square {\mathcal G}_2$ with strong  spatial-temporal
correlation, see Theorems \ref{gft1approximation.thm} and \ref{gft2approximation.thm}.  
In this paper, we also show that the proposed  GFTs ${\mathcal F}_\square$ and ${\mathcal F}_\otimes$ coincide only in the undirected graph setting, which
become essentially the joint GFT in \cite{Loukas16, Grassi18, Jiang2021}, see
Theorem \ref{sameGFT.thm}.
 The computational complexity to define the GFTs ${\mathcal F}_\square$ and ${\mathcal F}_\otimes$ are about $O(N_1^3N_2^3)$ and  $O(N_1^3+N_2^3)$,
 where 
 we say that $A=O(B)$ for two positive quantities $A$ and $B$ if $A/B$ is bounded by some absolute constant.
 In Section \ref{simulations.section}, we demonstrate the effectiveness of the proposed GFTs on  denoising the  
 hourly temperature data set collected at 32 weather stations
 in the region of Brest (France) in
 January 2014, and the one collected at 218 locations in the United States on August 1st, 2010.
     All proofs are collected in the  Appendix. 

{\bf Notation}: We use boldface upper and lower letters to represent matrices and vectors,
and ${\bf I}_N, {\bf O}_N$ and ${\bf 0}_N$ to denote the identity matrix, zero matrix  and zero column vector of  size $N$ respectively. We denote
the Euclidean norm of a vector ${\bf x}$ by $\|{\bf x}\|_2$,
 the transpose, vectorization and Frobenius norm of a matrix ${\bf A}$ by ${\bf A}^T$, ${\rm vec}({\bf A})$ and
$\|{\bf A}\|_F$ respectively,
and  the Kronecker product of two matrices ${\bf A}$ and ${\bf B}$ by ${\bf A}\otimes {\bf B}$.

%

\section{GFT 
 on  directed Cartesian product  graphs}  
\label{dft.section1}

Let ${\mathcal G}_1=(V_1, E_1)$ and ${\mathcal G}_2=(V_2, E_2)$ be two directed graphs of orders $N_1$ and $N_2$.  Our illustrative example
is the temporal line graph ${\mathcal T}$ and  spatial graph ${\mathcal S}$  to describe
 time-varying data sets on directed networks.
The
{\em Cartesian product graph}  $\mathcal{G}:={\mathcal G}_1\square {\mathcal G}_2=(V_1\times V_2, E_1\square E_2)$ has
vertices $(v_1, v_2)\in V_1\times V_2$ and edges between vertices
$(v_1 , v_2)$ and $(\tilde v_1 , \tilde v_2)$  if either $(v_1,\tilde v_1)\in E_1$ and $\tilde v_2=v_2$, or
$\tilde v_1=v_1$ and $(v_2, \tilde v_2)\in E_2$  
\cite{Loukas16, Grassi18, Jiang2021, ncjs22, Jiang22}.
In this section, following the approach in \cite{Yang2022},  we introduce a GFT ${\mathcal F}_\square$ on the directed
Cartesian product graph ${\mathcal G}$ and show that  graph signals 
with strong spatial-temporal correlation may have their energy
mainly concentrated on the low frequencies of the proposed GFT ${\mathcal F}_\square$,
 see Theorem \ref{gft1approximation.thm}.

Denote the adjacency, in-degree and
(in-degree) Laplacian matrices of graphs ${\mathcal G}_l$ by
${\bf A}_l, {\bf D}_l$ and ${\bf L}_l={\bf D}_l-{\bf A}_l, l=1, 2$, respectively. One may verify that
the adjacency
and
 Laplacian matrices of the Cartesian product graph ${\mathcal G}$ are given by
\begin{equation*}
{\bf A}_{\square}={\bf A}_{1}\otimes {\bf I}_{N_2} + {\bf I}_{N_1}\otimes {\bf A}_{2}
\end{equation*}
and
\begin{equation}\label{lap.def}
{\bf L}_{\square}={\bf L}_{1}\otimes {\bf I}_{N_2} + {\bf I}_{N_1}\otimes {\bf L}_{2}.
\end{equation}

%

%
%
%

A  signal on the Cartesian product graph ${\mathcal G}$
is usually represented by a matrix
${\bf X}=[{\bf x}_i]_{i\in V_1}\in\mathbb{R}^{N_1\times N_2}$ and its vectorization
${\bf x}={\rm vec}({\bf X})$, where for every $i\in V_1$, ${\bf x}_i$ is a graph signal on ${\mathcal G}_2$.
It could also be  represented by a matrix ${\bf Y}=[ {\bf y}_j^T]_{j\in V_2}$ and its vectorization ${\bf y}={\rm vec}({\bf Y})$, where
for every $j\in V_2$, ${\bf y}_j$ is a graph signal on ${\mathcal G}_1$.
For our illustrative  temporal-spatial (time-varying) scenario,
 ${\bf x}_i$ is the spatial signal at time $i\in V_1$ and ${\bf y}_j$ is the temporal signal at the vertex $j\in V_2$.

For the  Laplacian  $\L_\square$ on the directed Cartesian product graph ${\mathcal G}$, we take its SVD
as follows,
\begin{equation}\label{svd.def}
\L_\square=\U\pmb \Sigma\V^T=\sum_{k=0}^{N-1} \sigma_k {\bf u}_k {\bf  v}_k^T, \end{equation}
 where $N=N_1N_2$,
$ {\bf U}=[{\bf u}_0, \ldots, {\bf u}_{N-1}]$ and ${\bf V}=[{\bf v}_0, \ldots, {\bf v}_{N-1}]$
 are  orthogonal matrices,
  and the diagonal matrix
 $\pmb \Sigma={\rm diag} (\sigma_0, \ldots, \sigma_{N-1})$
   has   singular values of the  Laplacian $\L_\square$  deployed on the diagonal
   in a nondecreasing order, i.e.,
$$    0=\sigma_0\le\sigma_1\le\ldots\le\sigma_{N-1}.$$
The computational complexity to perform the  SVD in \eqref{svd.def}  is $O(N^3)$
\cite{lloyd1997}.
 For the undirected graph setting, i.e., ${\mathcal G}_1$ and ${\mathcal G}_2$ are undirected graphs,
the  Laplacian matrices  ${\bf L}_l, l=1, 2$, are positive semi-definite and they have the following eigen-decomposition
\begin{equation}\label{eigendecomposition.def2}
{\bf L}_l=\sum_{i=0}^{N_l-1} \lambda_{l, i} {\bf w}_{l, i} {\bf w}_{l, i}^T,\  l=1, 2,
\end{equation}
where $0=\lambda_{l,0}\le \ldots\le \lambda_{l, N_l-1}$
 are eigenvalues of ${\bf L}_l$,
and ${\bf w}_{l i}, 0\le i\le N_l-1$,  form an orthonormal basis of ${\mathbb R}^{N_l}$.
Therefore singular values of the   Laplacian  $\L_\square$ on the undirected Cartesian product graph ${\mathcal G}$
are  the sum of eigenvalues of ${\bf L}_1$ and ${\bf L}_2$, and
orthogonal matrices ${\bf U}$ and ${\bf V}$ are the same and consist  of  Kronecker products of eigenvectors of Laplacians
${\bf L}_1$ and ${\bf L}_2$, i.e.,
\begin{eqnarray}\label{undirectedlap.eig}
\L_\square & \hskip-0.08in =& \hskip-0.08in  \sum_{i=0}^{N_1-1} \sum_{j=0}^{N_2-1}
(\lambda_{1, i}+\lambda_{2, j})
({\bf w}_{1, i}\otimes {\bf w}_{2, j}) ({\bf w}_{1, i}\otimes {\bf w}_{2, j})^T
\end{eqnarray}
\cite{Grassi18, Chungbook,  Merris94}. 
This implies that  the computational complexity to perform the SVD \eqref{undirectedlap.eig} (and also the eigen-decomposition) of  the  Laplacian  ${\bf L}_\square$
in the undirected graph setting is $O(N_1^3+N_2^3)$ \cite{lloyd1997}, instead of
$O(N_1^3N_2^3)$ in the general directed graph setting.

Based on the SVD \eqref{svd.def}
 of the  Laplacian matrix $\L_\square$, we can follow the approach in \cite{Yang2022} to define the
 GFT  on the directed Cartesian product graph ${\mathcal G}$. 

\begin{definition}\label{jgft1.def}
{\rm Let $\cal G$ be the Cartesian product  of directed graphs ${\mathcal G}_1$ and ${\mathcal G}_2$,   the  Laplacian $\L_\square$  on ${\mathcal G}$ be given in \eqref{lap.def}, and orthogonal matrices  $\U, \V$  of size $N\times N$ be as in \eqref{svd.def}.   We define the {\em graph Fourier transform}
${\mathcal F}_\square: {\mathbb R}^{N}\longmapsto {\mathbb R}^{2N}$    on 
${\mathcal G}$  by
 \begin{eqnarray}\label{jointfourier1.def}
{\mathcal F}_\square{\bf x} & \hskip-0.08in := & \hskip-0.08in
\frac{1}{2}\begin{pmatrix}
 ({\bf U}+{\bf V})^T{\bf x}\\
 ({\bf U}-{\bf V})^T{\bf x}
\end{pmatrix}
=  \hskip 0.08in
 \frac{1}{2}
\begin{pmatrix}
({\bf u}_0+{\bf v}_0)^T{\bf x}\\
\vdots\\
({\bf u}_{N-1}+{\bf v}_{N-1})^T{\bf x}\\
({\bf u}_0-{\bf v}_0)^T{\bf x}\\
\vdots\\
({\bf u}_{N-1}-{\bf v}_{N-1})^T{\bf x}\\
\end{pmatrix},
\end{eqnarray}
where  $\bf x$ is a graph signal on the Cartesian product graph  $\cal G$.
 We also define
 the  {\em inverse  graph Fourier transform} ${\mathcal F}_\square^{-1}: {\mathbb R}^{2N}\longmapsto {\mathbb R}^{N}$ by
 \begin{eqnarray}\label{inverFT1.def}
& \hskip-0.08in & \hskip-0.08in {\mathcal F}_\square^{-1}\begin{pmatrix}
{\bf z}_1\\
{\bf z}_2
\end{pmatrix} 
 :=\frac{1}{2} \big({\bf U} ({\bf z}_1+{\bf z}_2)+ {\bf V}({\bf z}_1-{\bf z}_2)\big)  
\nonumber \\
& \hskip-0.08in = & \hskip-0.08in  \frac{1}{2} \sum_{k=0}^{N-1}
({ z}_{1, k}+{ z}_{2, k}) {\bf u}_k+ ({z}_{1, k}-{z}_{2, k}) {\bf v}_k
\end{eqnarray}
 for ${\bf z}_l=[z_{l, 0}, \ldots, z_{l,N-1}]^T\in \R^{N}, l=1, 2$.
}
\end{definition}

 For the proposed  GFT  ${\mathcal F}_\square$ and  inverse GFT  ${\mathcal F}_\square^{-1}$ in Definition \ref{jgft1.def}, one may verify that
  \begin{equation}\label{inversegft.eqn1}
 {\mathcal F}_\square^{-1}{\mathcal F}_\square {\bf x}={\bf x},
 \end{equation}
where ${\bf x}$  is a  signal on the Cartesian product graph ${\mathcal G}$.
From the orthogonal properties of matrices ${\bf U}$ and ${\bf V}$
it follows that Parseval's identity
\begin{equation}
\|{\mathcal F}_\square{\bf x}\|_2= \|{\bf x}\|_2
\end{equation}
hold for all  signals ${\bf x}$ on
${\mathcal G}$.

Following the terminology in \cite{Yang2022},
we may  use singular values $\sigma_k, 0\le k
\le N-1$, as {\em frequencies} of the proposed GFT ${\mathcal F}_\square$,
and   ${\bf u}_k, {\bf v}_k, 0\le k\le N-1$,  as the associated {\em left/right frequency components}.
In the following theorem, we show that signals
on the directed Cartesian product graph ${\mathcal G}$ with strong spatial-temporal correlation may have their  energy mainly concentrated on the low frequencies of
the proposed GFT ${\mathcal F}_\square$, see
Appendix \ref{proof2.appendix} for the proof.

\begin{theorem}
\label{gft1approximation.thm}
Let $\cal G$ be the Cartesian product of directed graphs ${\mathcal G}_1$ and ${\mathcal G}_2$,  $\L_\square$  be the  Laplacian  \eqref{lap.def} on ${\mathcal G}$, and ${\bf u}_k, {\bf v}_k, \sigma_k, 0\le k\le N-1$ be  as in \eqref{svd.def}, where $N=N_1N_2$,
$N_1$ and $N_2$ are the orders of graphs ${\cal G}_1$ and  ${\cal G}_2$.
For a frequency bandwidth  $M\in \{1, 2, \ldots, N\}$, define the low frequency component of a graph signal $\bf x$ on ${\mathcal G}$ with bandwidth $M$ by
\begin{eqnarray}\label{gft1approximation.thm.eq1}
{\bf x}_{M,\square} & \hskip-0.08in = & \hskip-0.08in \frac{1}{2} \sum_{k=0}^{M-1} (z_{1, k}+z_{2, k}){\bf u}_k+ (z_{1, k}-z_{2, k}){\bf v}_k\nonumber\\
&\hskip-0.08in = & \hskip-0.08in \frac{1}{2}  \sum_{k=0}^{M-1}({\bf u}_k{\bf u}_k^T+ {\bf v}_k{\bf v}_k^T){\bf x},
\end{eqnarray}
where $z_{1, k}= ({\bf u}_k+{\bf v}_k)^T {\bf x}/2$ and $z_{2, k}= ({\bf u}_k-{\bf v}_k)^T {\bf x}/2, 0\le k\le M-1$. Then
\begin{eqnarray}
\hskip-0.08in & \hskip-0.08in & \|{\bf x}-{\bf x}_{M,\square}\|_2  \le   \frac{1}{2 \sigma_{M-1}}
\big(\| {\bf L}_\square {\bf x}\|_2+ \|{\bf L}_\square^T{\bf  x}\|_2\big)\nonumber\\
\hskip-0.08in & \le \hskip-0.08in &   \frac{1}{2 \sigma_{M-1}}
\big( \| ({\bf L}_1\otimes {\bf I}_{N_2}) {\bf x}\|_2+ \|({\bf L}_1^T\otimes {\bf I}_{N_2}) {\bf  x}\|_2\nonumber\\
\hskip-0.18in & \hskip-0.08in  &  +   \hskip0.05in
\| ({\bf I}_{N_1}\otimes {\bf L}_2) {\bf x}\|_2+ \|({\bf I}_{N_1}\otimes {\bf L}_2^T) {\bf  x}\|_2\big),
\end{eqnarray}
where $\sigma_{M-1}$ is the cut-off  frequency of the bandlimiting procedure \eqref{gft1approximation.thm.eq1}.
\end{theorem}

\section{GFT 
on directed product graphs}
\label{novelgft.section}

Graph signals in some applications, such as time-varying signals, carry different correlation characteristics in  different directions, and hence
GFT in such scenario should be designed to reflect spectral characteristic for different directions
\cite{
Grassi18, ncjs22, Jiang22, Kurokawa17}. 
 In this section, based on the SVDs of Laplacians on ${\mathcal G}_1$ and ${\mathcal G}_2$,
  we introduce another GFT ${\mathcal F}_\otimes$ on the product graph ${\mathcal G}$, see Definition \ref{jgft2.def}. Comparing with the GFT ${\mathcal F}_\square$ in Definition \ref{jgft1.def}, the new GFT ${\mathcal F}_\otimes$ has
   lower computational complexity. On the other hand,
  they have similar performance to efficiently represent
  time-varying signals with strong correlation,  see Theorem \ref{gft2approximation.thm} and  numerical demonstrations in Section \ref{simulations.section}. 
In this section, we also  show that the proposed GFTs ${\mathcal F}_\otimes$ and ${\mathcal F}_\square$ coincide only in the undirected graph setting,
see Theorem \ref{sameGFT.thm}.

Let ${\mathcal G}_1=(V_1, E_1)$ and ${\mathcal G}_2=(V_2, E_2)$ be two directed graphs,
and denote their Laplacians and orders by ${\bf L}_l$ and $N_l, l=1, 2$ respectively.
 For  the Laplacian matrices  ${\bf L}_l, l=1, 2$, we take their SVDs
\begin{equation}\label{Vsvd.def}
\L_{l}=\U_l{\pmb \Sigma}_l\V_l^T=\sum_{i=0}^{N_l-1} \sigma_{l,i} {\bf u}_{l, i} {\bf v}_{l, i}^T,\end{equation}
where $\sigma_{l,i}, 0\le i\le N_l-1$,  are   singular values of the Laplacian matrix $\L_l$
with a nondecreasing order, $\U_l=[{\bf u}_{l,0},\ldots, {\bf u}_{l, N_l-1}]$ and $\V_l=[{\bf v}_{l,0},\ldots, {\bf v}_{l, N_l-1}]$
are orthonormal matrices. 
Set
\begin{equation}\label{orthogonalMatrix2.def}
\U_\otimes = \U_1\otimes\U_2\  \ \mbox{and}\  \  \V_\otimes = \V_1\otimes\V_2.
\end{equation}
With the help of SVDs of Laplacians ${\bf L}_l, 1=1, 2$, we propose  the second GFT on the directed product graph ${\mathcal G}$ as follows.

\begin{definition}\label{jgft2.def}{\rm
Let directed graphs $\G_l, l\in \{1, 2\}$, have  orders $N_l$ and  Laplacian matrices  $\L_l$, orthogonal matrices $\U_l, \V_l$ be given as in \eqref{Vsvd.def},
  $\U_\otimes$ and $\V_\otimes$ be the orthogonal matrices in \eqref{orthogonalMatrix2.def}, and set $N=N_1N_2$. Then we define the   {\em  graph Fourier transform}
${\mathcal F}_\otimes: {\mathbb R}^{N}\longmapsto {\mathbb R}^{2N}$
and {\em inverse graph Fourier transform} ${\mathcal F}_\otimes^{-1}: {\mathbb R}^{2N}\longmapsto {\mathbb R}^{N}$
  on the   product graph ${\mathcal G}$ by
 \begin{equation}\label{jgft2.def.eq1}
{\mathcal F}_\otimes{\bf x} :=
\frac{1}{2} \left(\begin{array}{c} 
 ({\bf U}_\otimes+{\bf V}_\otimes)^T{\bf x}\\
  ({\bf U}_\otimes-{\bf V}_\otimes)^T{\bf x}
 \end{array}\right) 
\end{equation}
and
\begin{equation}\label{inverFT2.def}
{\mathcal F}_\otimes^{-1}\begin{pmatrix}
{\bf z}_1\\
{\bf z}_2
\end{pmatrix} 
:=\frac12 \big(\U_\otimes({\bf z}_1+{\bf z}_2)+\V_\otimes({\bf z}_1-{\bf z}_2)\big),
\end{equation}
 where  ${\bf x}\in \R^{N}$ is a  signal on the graph ${\mathcal G}$, and  ${\bf z}_1, {\bf z}_2$ are vectors in $\R^{N}$.
}\end{definition}

For the  GFT ${\mathcal F}_\otimes$ just defined, 
 we may use  pairs
 $(\sigma_{1, i}, \sigma_{2, j})$ of singular values of Laplacians  ${\bf L}_1$ and ${\bf L}_2$ as {\em frequency pairs} of the proposed GFT,
and   ${\bf u}_{1,i}\otimes {\bf u}_{2, j}$ and ${\bf v}_{1, i}\otimes {\bf v}_{2, j}, 0\le i\le N_1-1, 0\le j\le N_2-1$, as the associated {\em left/right frequency components}.
The computational complexity to evaluate the  left/right frequency components of the
GFT ${\mathcal F}_\otimes$ is $O(N_1^3+N_2^3)$ \cite{lloyd1997}, c.f.
$O(N_1^3+N_2^3)$ to evaluate the left/right frequency components
${\bf u}_k, {\bf v}_k, 0\le k\le N_1N_2-1$, of the GFT ${\mathcal F}_\square$
\eqref{jointfourier1.def} in the undirected graph setting, and $O(N_1^3N_2^3)$ to evaluate them in general directed graph setting, see \eqref{undirectedlap.eig} and \eqref{svd.def}
and also numerical simulations in Section \ref{simulations.section}.

By the orthogonality of the matrices $\U_l, \V_l, l=1,2$, 
one may verify that
\begin{equation}
\|{\mathcal F}_\otimes{\bf x}\|_2= \|{\bf x}\|_2
\end{equation}
and
\begin{equation}\label{inverse2gft.eq}
{\mathcal F}_\otimes^{-1}{\mathcal F}_\otimes {\bf x}={\bf x}
\end{equation}
hold for all  signals ${\bf x}$ on the  product graph ${\mathcal G}$.
Similar to the conclusion in Theorem \ref{gft1approximation.thm}, we can show that bandlimiting in
the frequency domain of the GFT ${\mathcal F}_\otimes$ provides  good approximations to graph signals
with strong  spatial-temporal correlation, 
 see
Appendix \ref{proof3.appendix} for the proof.

\begin{theorem}
\label{gft2approximation.thm}
Let $\cal G$ be the Cartesian product  of directed graphs ${\mathcal G}_1$ and ${\mathcal G}_2$,
 $\sigma_{l,i},  {\bf u}_{l, i},  {\bf v}_{l, i}, 0\le i\le N_l-1, l=1,2$, be as in \eqref{Vsvd.def},
and $\mu_k, 0\le k\le N-1$, be the
ascending order of  $\sigma_{1, i}+\sigma_{2, j}, 0\le i\le N_1-1, 0\le j\le N_2-1$, where $N=N_1N_2$.
For a frequency bandwidth $1\le M\le N$ of the GFT ${\mathcal G}_\otimes$ in \eqref{jgft2.def.eq1}, define the low frequency component of a graph signal $\bf x$ on ${\mathcal G}$ with bandwidth $M$ by
\begin{eqnarray}\label{gft2approximation.thm.eq1}
{\bf x}_{M,\otimes}& \hskip-0.08in = & \hskip-0.08in\frac{1}{2}\sum_{(i,j)\in {\mathcal S}_M}({\bf u}_{1,i}\otimes {\bf u}_{2,j})({\bf u}_{1,i}\otimes {\bf u}_{2,j})^T{\bf x}\nonumber\\
& \hskip-0.08in &  +({\bf v}_{1,i}\otimes {\bf v}_{2,j})({\bf v}_{1,i}\otimes {\bf v}_{2,j})^T{\bf x},
\end{eqnarray}
where ${\mathcal S}_M$ contains  all pairs $(i,j)$ with $\sigma_{1,i}+\sigma_{2,j}$ being some $\mu_k, 0\le k\le M-1$.
 Then
\begin{eqnarray} \label{gft2approximation.thm.eq2}
\|{\bf x}-{\bf x}_{M,\otimes}\|_2 \hskip-0.08in  & \le \hskip-0.08in  &    \frac{1}{ 2 \mu_{M-1}}
\big( \| ({\bf L}_1\otimes {\bf I}_{N_2}) {\bf x}\|_2\nonumber\\
\hskip-0.10in & \hskip-0.08in  & \hskip-0.08in  + \|({\bf L}_1^T\otimes {\bf I}_{N_2}) {\bf  x}\|_2
 +
\| ({\bf I}_{N_1}\otimes {\bf L}_2) {\bf x}\|_2
 + \|({\bf I}_{N_1}\otimes {\bf L}_2^T) {\bf  x}\|_2\big),
%
\end{eqnarray}
where $\mu_{M-1}$ is the cut-off  frequency of the bandlimiting procedure  \eqref{gft2approximation.thm.eq1}.
\end{theorem}


For a graph signal ${\bf X}=[{\bf x}_i]_{i\in V_1}\in\mathbb{R}^{N}$ or its vectorization
${\bf x}={\rm vec}({\bf X})$ on the product graph ${\mathcal G}$, using
the mixed Kronecker matrix-vector product  property, we can rewrite
 its GFT  ${\mathcal F}_\otimes {\bf x}$ as follows:
\begin{equation}
{\mathcal F}_\otimes {\bf x}
=\frac{1}{2}
 \left(\begin{array}{c} 
{\rm vec} \big( {\bf U}_2^T {\bf X}{\bf U}_1+ {\bf V}_2^T {\bf X}{\bf V}_1\big)\\
{\rm vec} \big( {\bf U}_2^T {\bf X}{\bf U}_1- {\bf V}_2^T {\bf X}{\bf V}_1\big)
 \end{array}\right).
\end{equation}
Thus just as taking classical discrete Fourier transform of two-dimensional signals by directions, we can implement the GFT  ${\mathcal F}_\otimes {\bf x}$
 in the direction of the graph ${\mathcal G}_1$ and then of the graph ${\mathcal G}_2$, or vice versa, see Algorithm \ref{fourier.algorithm}.

\begin{algorithm}[ht]
\caption{Algorithm to implement the GFT  ${\mathcal F}_\otimes$}
\label{fourier.algorithm}
\begin{algorithmic} 

\STATE {\bf Input}:  Graph  signal ${\bf X}$.

\STATE{\bf Steps}:
\STATE{\bf 1)} Do $ {\bf Y}_1= {\bf X}{\bf U}_1$ and $\widetilde {\bf Y}_1={\bf X}{\bf V}_1$;\\
\STATE{\bf 2)}
 Do ${\bf Y}_2={\bf U}_2^T{\bf Y}_1$ and $\widetilde {\bf Y}_2={\bf V}_2^T\widetilde {\bf Y}_1$;\\
\STATE{\bf 3)}
Do $\widehat {\bf X}_1= ({\bf Y}_2+\widetilde {\bf Y}_2)/2$ and $\widehat {\bf X}_2= ({\bf Y}_2-\widetilde {\bf Y}_2)/2$.\\

\STATE {\bf Outputs}:
 The first component $\widehat {\bf X}_1$ and the second component $\widehat {\bf X}_2$ of the GFT ${\mathcal F}_\otimes {\rm vec}({\bf X})$.

\end{algorithmic}
\end{algorithm}

Similarly, we have
$$ {\mathcal F}_\otimes^{-1}\begin{pmatrix}
{\bf z}_1\\
{\bf z}_2
\end{pmatrix}=\frac{1}{2}\big( {\bf U}_2 ({\bf Z}_1+{\bf Z}_2) {\bf U}_1^T+
{\bf V}_2 ({\bf Z}_1-{\bf Z}_2) {\bf V}_1^T\big)
$$
 for ${\bf z}_1, {\bf z}_2\in {\mathbb R}^{N}$, where ${\bf Z}_1={\rm vec}^{-1}({\bf z}_1)$ and ${\bf Z}_2={\rm vec}^{-1}({\bf z}_2)$, see Algorithm
\ref{inversefourier.algorithm} for the implementation.

\begin{algorithm}[ht]
\caption{Algorithm to implement the inverse GFT  ${\mathcal F}_\otimes^{-1}$}
\label{inversefourier.algorithm}
\begin{algorithmic} 

\STATE {\bf Inputs}:  ${\bf z}_1, {\bf z}_2\in {\mathbb R}^N$.

\STATE {\bf Inverse vectorization}: ${\bf Z}_1={\rm vec}^{-1}({\bf z}_1)$ and ${\bf Z}_2={\rm vec}^{-1}({\bf z}_2)$.
\STATE{\bf Steps}:

\STATE{\bf 1)} Do $ {\bf W}_1= ({\bf Z}_1+{\bf Z}_2){\bf U}_1^T$ and $\widetilde {\bf W}_1=({\bf Z}_1-{\bf Z}_2){\bf V}_1^T$;\\
\STATE{\bf 2)}
 Do ${\bf W}_2={\bf U}_2{\bf W}_1$ and $\widetilde {\bf W}_2={\bf V}_2\widetilde {\bf W}_1$;\\
\STATE{\bf 3)}
 Do $ {\bf X}= ({\bf W}_2+\widetilde {\bf W}_2)/2$.\\

\STATE {\bf Output}:
 ${\bf x}={\rm vec}(  {\bf X}) = {\mathcal F}_\otimes^{-1}\begin{pmatrix}
{\bf z}_1\\
{\bf z}_2
\end{pmatrix}$.

\end{algorithmic}
\end{algorithm}

In the undirected  graph setting,  we obtain from
\eqref{eigendecomposition.def2} that
${\bf U}, {\bf V}$ in \eqref{svd.def} and
${\bf U}_\otimes, {\bf V}_\otimes$
in \eqref{orthogonalMatrix2.def} can be chosen to be the same, i.e.,
${\bf U}= {\bf V}= {\bf U}_\otimes= {\bf V}_\otimes$.
Therefore
\begin{equation}\label{undirect.gft} {\mathcal F}_\square{\bf x}={\mathcal F}_\otimes {\bf x}=
 \left(\begin{array}{c} 
 {\bf U}_\otimes^T{\bf x}\\
 {\bf 0}_N
 \end{array}\right)
\end{equation}
hold for all  signals ${\bf x}$ on the Cartesian product  of two undirected graphs. 
We remark that in the undirected graph setting,  ${\bf U}_\otimes^T{\bf x}$ is used in
\cite{Loukas16, Grassi18, Jiang2021} to define the joint GFT of a graph signal ${\bf x}$ on the product graph.

In the following theorem, we show that the proposed  GFTs ${\mathcal F}_\square$ and ${\mathcal F}_\otimes$ coincide only in
the undirected graph setting, see Appendix \ref{proof.appendix} for the proof.

\begin{theorem}\label{sameGFT.thm} Let ${\mathcal F}_\square$ and ${\mathcal F}_\otimes$ be the GFTs on the Cartesian product  of two graphs ${\mathcal G}_1$ and ${\mathcal G}_2$.  Assume that ${\mathcal G}_1$ and ${\mathcal G}_2$ are not edgeless graphs.
If ${\mathcal F}_\square={\mathcal F}_\otimes$, then ${\mathcal G}_1$ and ${\mathcal G}_2$ are undirected graphs.
\end{theorem}

\section{Numerical simulations}
\label{simulations.section}



\begin{figure}
\centering
\includegraphics[width=40mm, height=36mm]{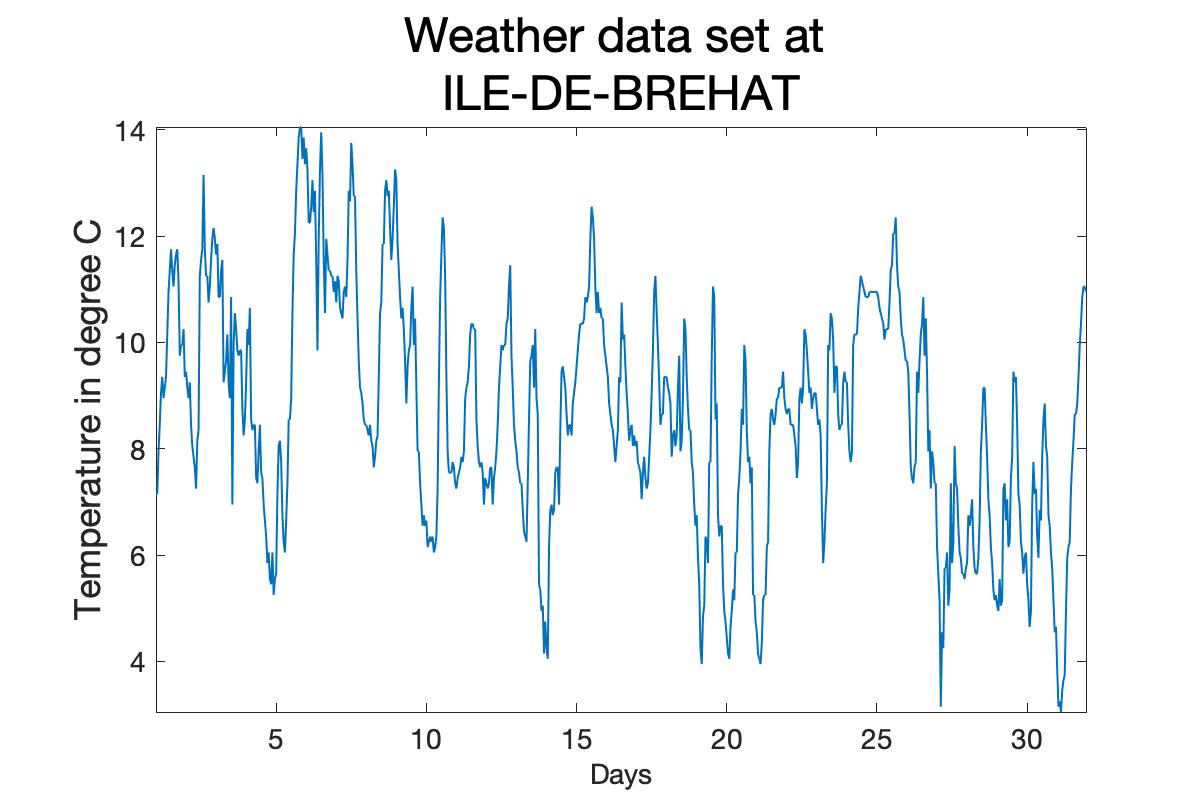}
\hskip .2in
\includegraphics[width=44mm, height=36mm]{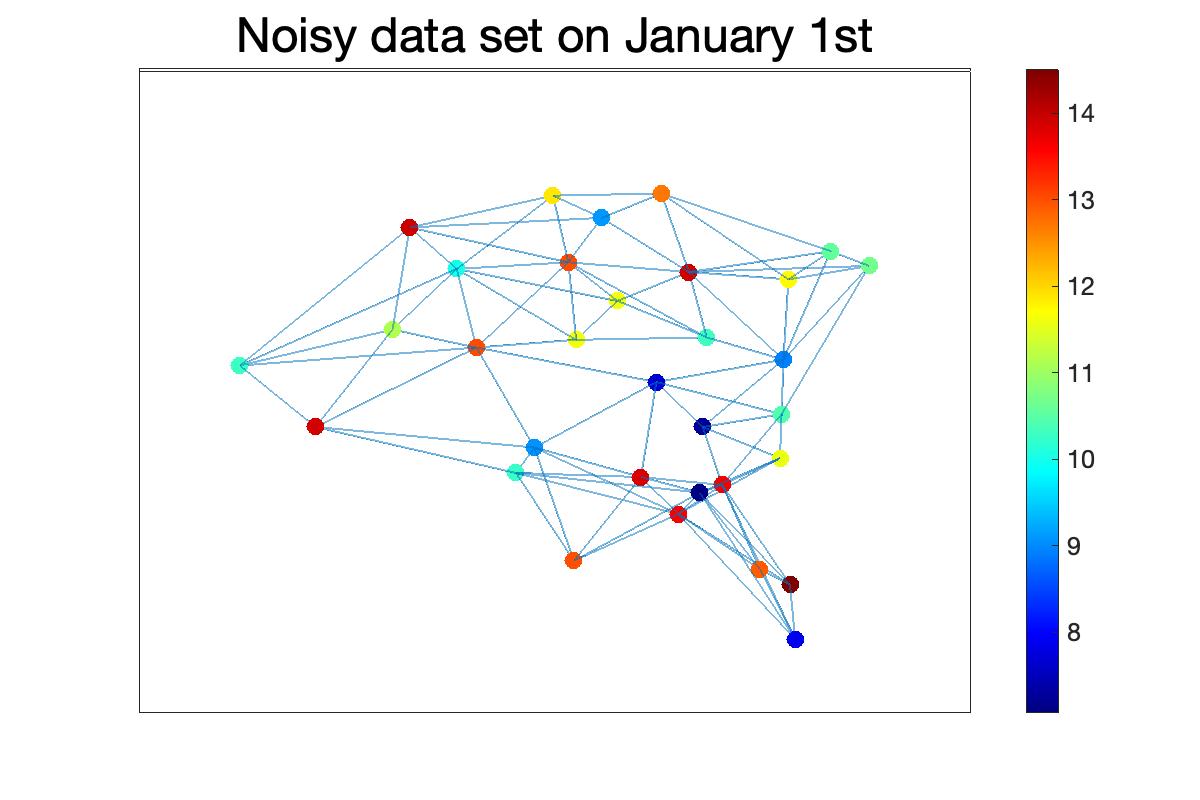}
\caption{
Plotted on the left is  the $24\times 31=744$ hourly temperature in Celsius
recorded  at the weather station  at the island of Brehat in January 2014, and on the right is
 the  noisy temperature data set $\tilde {\bf x}_1(12)={\bf x}_1(12)+{\pmb \eta}_1$
 at noon on January 1st, 2014, where
  entries of random noise vector ${\pmb \eta}_1$ are i.i.d. drawn on  $[-4, 4]$.
  }\label{denoise_M32.fig}
\end{figure}

In this section, we first consider the hourly temperature data set measured in Celsius
collected at 32 weather stations
 in the region of Brest (France) in
 January 2014, published by French national meteorological service \cite{stationary}. We represent the  above temperature data set by matrices  ${\bf X}_d=[{\bf x}_d(t_0)
  \ldots, {\bf x}_d({t_{23}})], 1\le d\le 31$, where the column vectors ${\bf x}_d(t_i), 0\le i\le 23$, are the regional temperature at $t_i$-th hour of  $d$-th day of
 January 2014.
 We consider matrices ${\bf X}_d, 1\le d\le 31$, as signals on the Cartesian product graph $\T \square {\mathcal S}$,
where ${\mathcal T}$ is the unweighted  directed line graph  with 24 vertices and
${\mathcal S}$ is the directed graph  with 32 locations of weather observation stations as vertices and edges constructed by the 5 nearest neighboring stations in physical distances,
and    weights on the edges are randomly chosen in $[0.8, 1.2]$, see Figure \ref{denoise_M32.fig}. 
In this section, we demonstrate the performances of the proposed  GFTs ${\mathcal F}_\square$ and ${\mathcal F}_\otimes$ by bandlimiting
the first $M$-frequencies of the noisy temperature data set
\begin{equation}
\label{noiseweather.eq}
\widetilde {\bf X}_d={\bf X}_d+ {\pmb \eta}_d,\  1\le d\le 31, \end{equation}
 where  ${\pmb \eta}_d$ are  additive  random noises  with entries being i.i.d. drawn on  $[-c, c]$ with $c\in [0, 8]$.
 All experiments are implemented on a Macbook pro (2.3 GHz quad-core Intel Core i7 and 32 GB memory)  by Matlab R2020b.

\begin{figure}
\centering
\includegraphics[width=68mm, height=42mm]{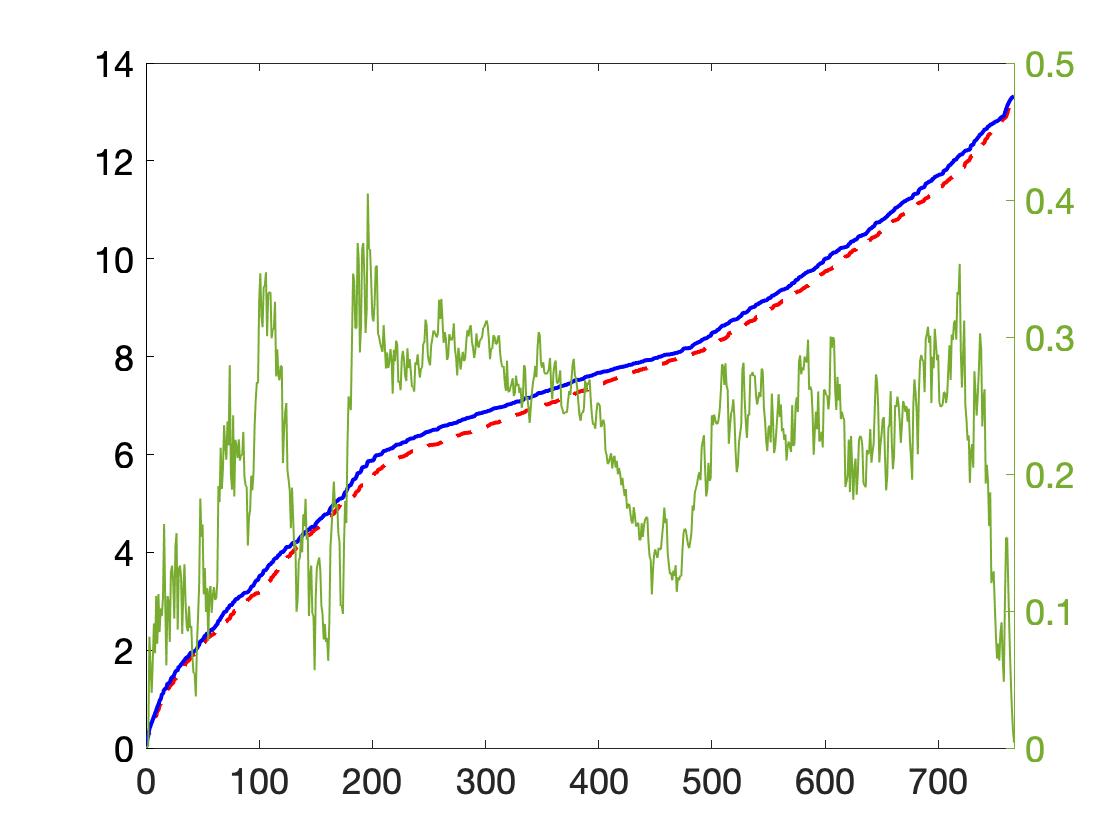}
\caption{
Plotted in red dotted line and  blue solid line are the frequencies $\sigma_k$ and $\mu_k, 0\le k\le 767$, associated with the GFTs $\cal F_\square$ and $\cal F_\otimes$  scaled at the left $y$-axis respectively. Plotted in lime green are the differences $\mu_k-\sigma_k, 0\le k\le 767$, scaled at the right $y$-axis.
  }
\label{freq.fig}
\end{figure}

For the unweighted directed line graph ${\mathcal T}$ of order $N_1=24$,  weighted directed graph ${\mathcal S}$  of order $N_2=32$ and their  Cartesian product graph $\T \square {\mathcal S}$ of order $768=24\times 32$, we take
 $ \sigma_k, {\bf u}_k, {\bf v}_k,0\le k\le 767$,  as in \eqref{svd.def},
and  $\sigma_{l, i}, {\bf u}_{l, i}, {\bf v}_{l, i},  0\le i\le N_l-1, l=1,2$. 
Inspired by the eigen-decomposition in \eqref{undirectedlap.eig}
and the coincidence \eqref{undirect.gft} of GFTs in the undirected graph setting, we arrange  frequencies
 of the GFT ${\mathcal F}_\square$ in the  ascending order $0=\sigma_0\le \ldots\le \sigma_{767}$,
and  frequency pairs $(\sigma_{1, i}, \sigma_{2, j})$ of the GFT ${\mathcal F}_\otimes$ in the ascending order of $\sigma_{1, i}+\sigma_{2, j}, 0\le i\le 23, 0\le j\le 31$, which are represented by
$0=\mu_0\le \ldots\le \mu_{767}$.
 The time to find the left/right frequency components
 ${\bf u}_k, {\bf v}_k,0\le k\le 767$, of the GFT ${\mathcal F}_\square$ and the ones
 ${\bf u}_{1, i}\otimes {\bf u}_{2, j}, {\bf v}_{1, i}\otimes {\bf v}_{2, j}, 0\le i\le 23, 0\le j\le 31$, of the GFT ${\mathcal F}_\otimes$
 are  0.0861 and 0.0189 seconds respectively.  This confirms that the GFT ${\mathcal F}_\otimes$
  has lower computational complexity than the GFT ${\mathcal F}_\square$ does.
Our numerical simulations also show that
 $0\le \sigma_k, \mu_k\le 13.3206$ and
$\sigma_k\le \mu_k\le \sigma_k+0.4047, 0\le k\le 767$, see Figure \ref{freq.fig}. 
  Therefore the proposed  GFTs ${\mathcal F}_\square$ and ${\mathcal F}_\otimes$ may have similar frequency information.


\begin{figure}
\centering
\includegraphics[width=38mm, height=32mm]{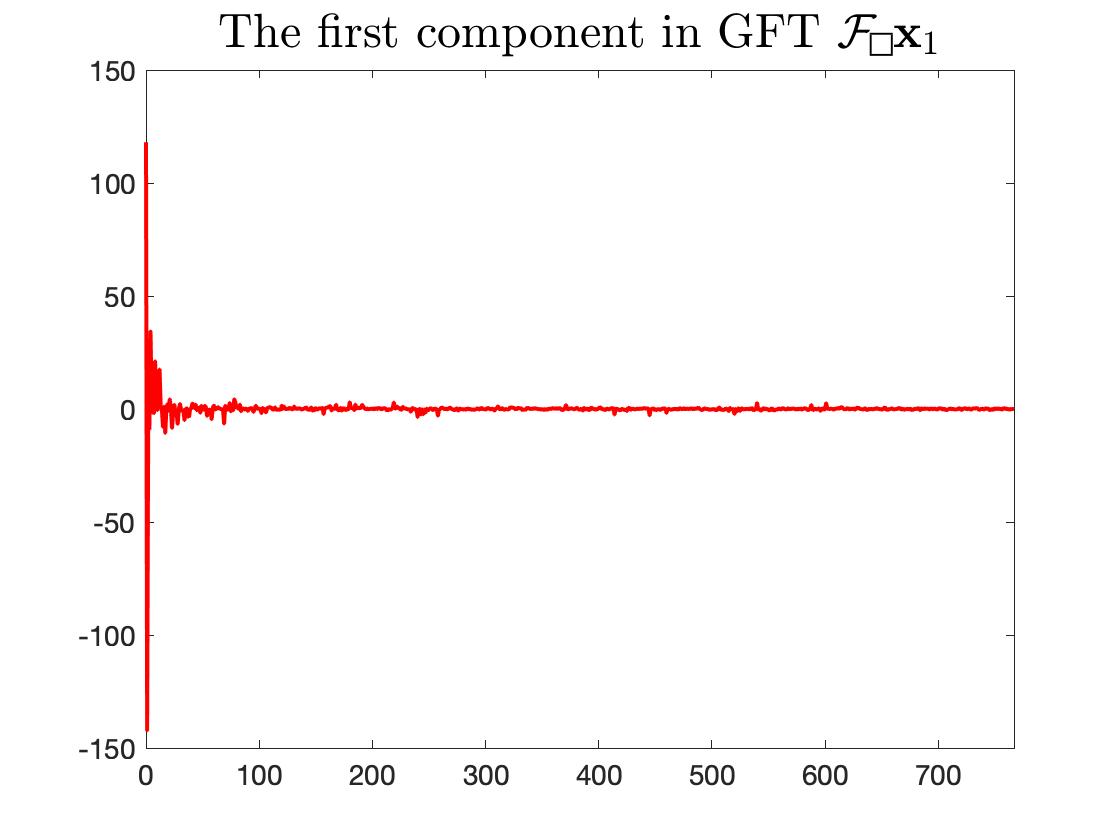}
\includegraphics[width=38mm, height=32mm]{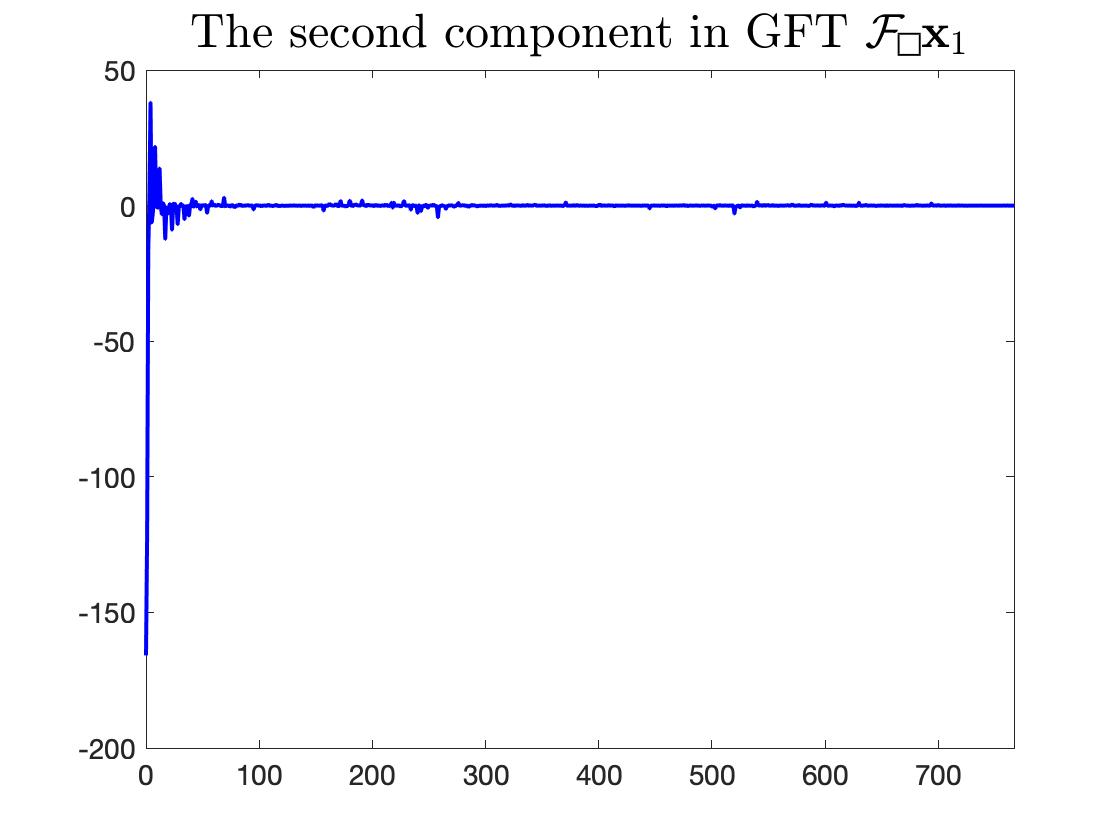}\\
\includegraphics[width=38mm, height=32mm]{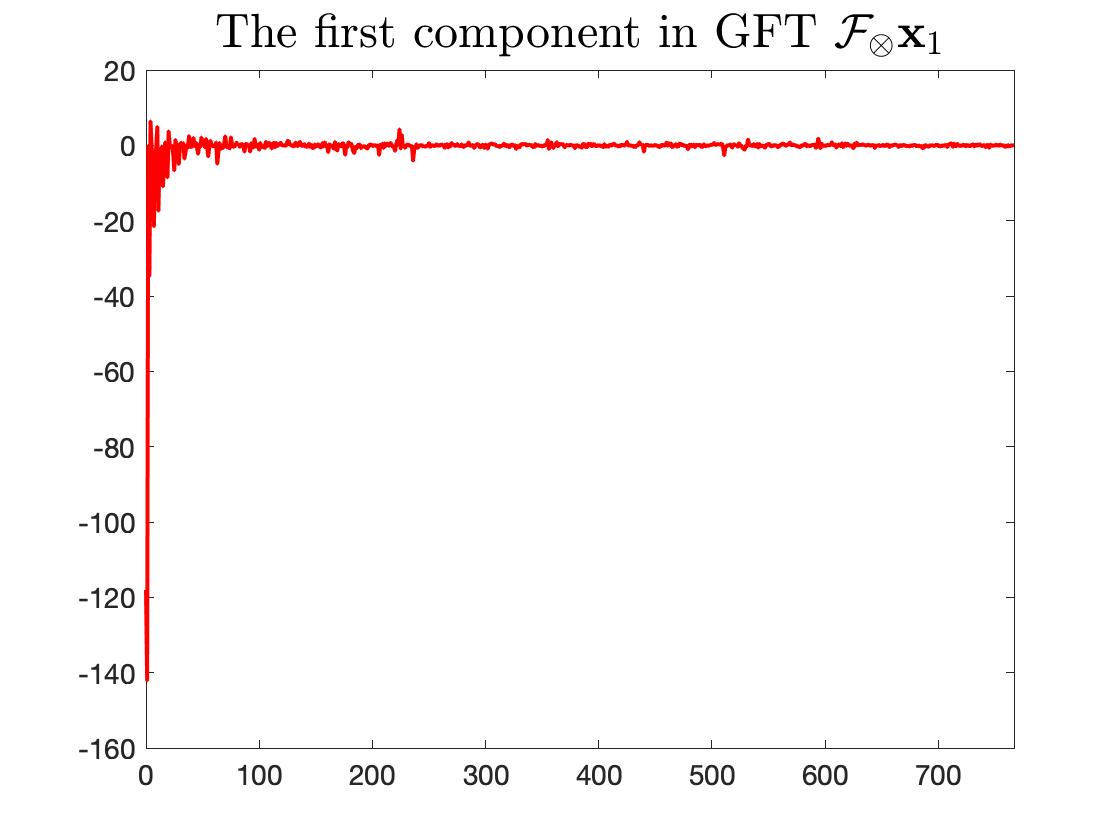}
\includegraphics[width=38mm, height=32mm]{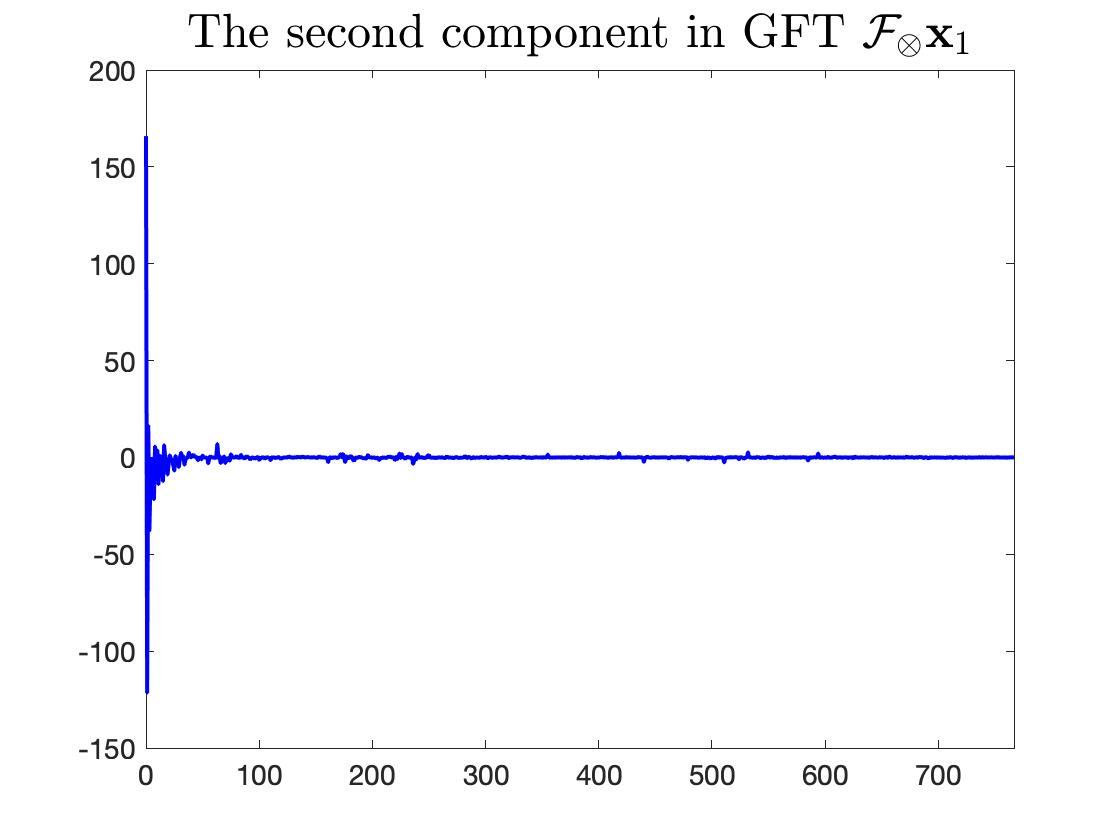}\\
\includegraphics[width=38mm, height=32mm]{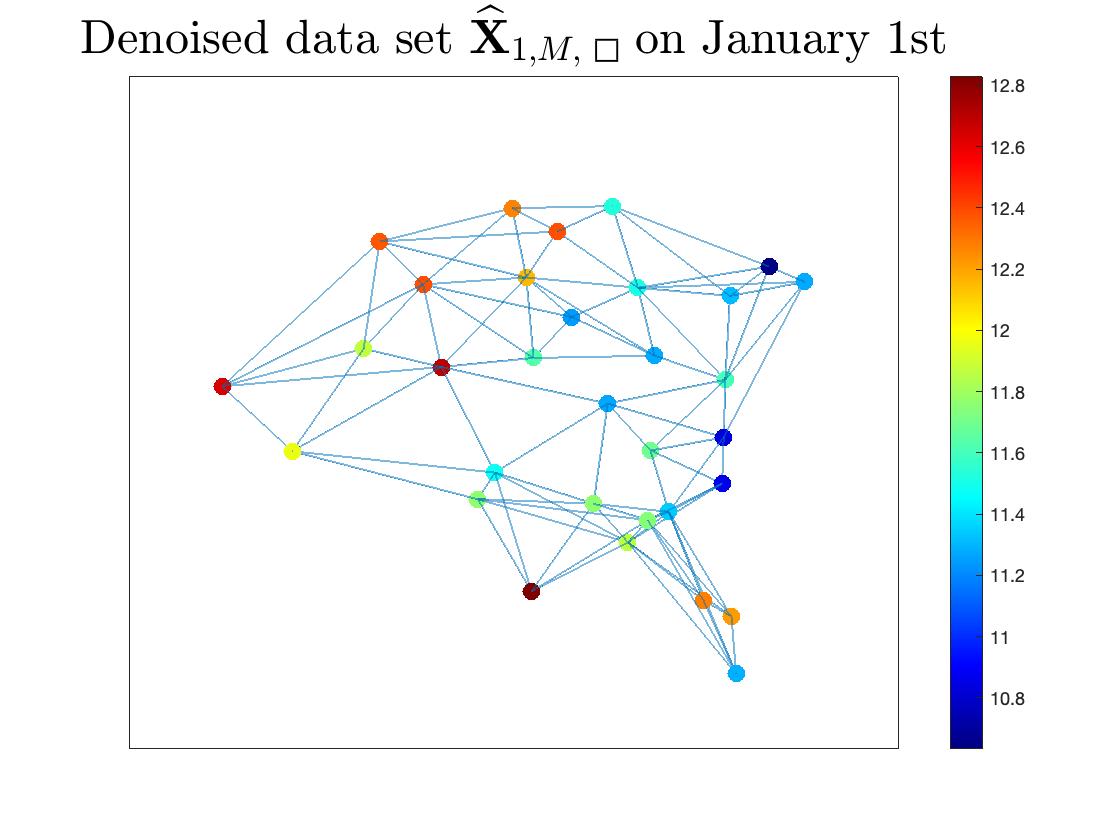}
\includegraphics[width=38mm, height=32mm]{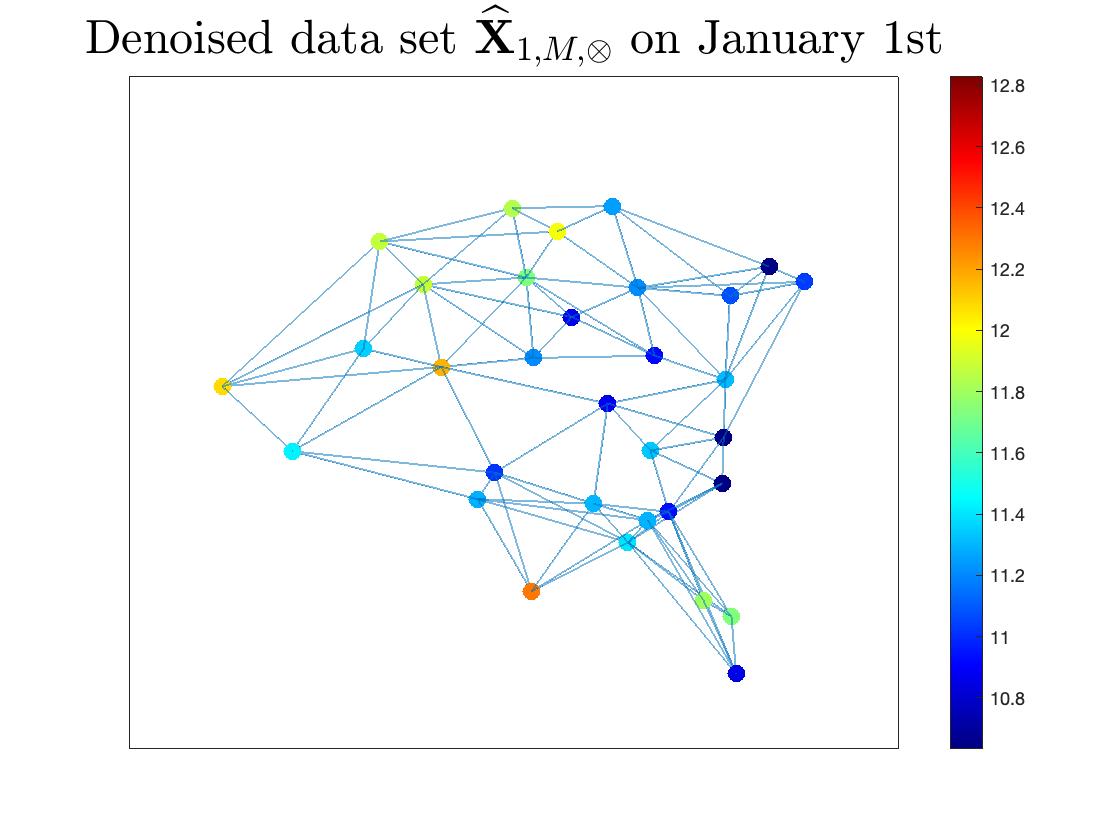}
\caption{
Plotted on the top left and right are the first component
  $({\bf U}+{\bf V})^T{\bf x}_1/2$ and the second component  $({\bf U}-{\bf V})^T{\bf x}_1/2$
  of the GFT ${\mathcal F}_\square {\bf x}_1$ of the signal ${\bf x}_1$ respectively,
  where $\U, \V$ are the orthogonal matrices in \eqref{svd.def} of $\L_\square$ and ${\bf x}_1={\rm vec}({\bf X}_1)$ is the vectorization of  the temperature data set on January 1st, 2014.
On the middle left and right are the first component
  $({\bf U}_\otimes+{\bf V}_\otimes)^T{\bf x}_1/2$ and the second component  $({\bf U}_\otimes-{\bf V}_\otimes)^T{\bf x}_1/2$
  of the GFT ${\mathcal F}_\otimes {\bf x}_1$ respectively,
  where $\U, \V$ are the orthogonal matrices in  \eqref{orthogonalMatrix2.def}.
 On the bottom left and right are  the snapshots of
 bandlimiting approximations $\widehat {\bf X}_{1, M, \square}$ and
 $\widehat {\bf X}_{1, M, \otimes}$ of the noisy temperature data set  $\widetilde {\bf X}_1$  at noon on  January 1st, 2014 
 respectively, where $c=4$ and $M=32$.}
\label{FourierGFT.fig}
\end{figure}

Let $1\le M\le 768$. Applying the  bandlimiting procedure of
the first $M$-frequencies of  the  GFTs ${\mathcal F}_\square$ and ${\mathcal F}_\otimes$
to
the noisy temperature data set
$\widetilde {\bf X}_d$ in \eqref{noiseweather.eq}, we obtain
\begin{equation}\label{square.simulation.eq1}
\widehat {\bf X}_{d, M, \square} 
=   {\rm vec}^{-1} \Big(\frac{1}{2}\sum_{k=0}^{M-1}\big ({\bf u}_k{\bf u}_k^T+{\bf v}_k {\bf v}_k^T\big) {\rm vec}(\widetilde {\bf X}_d)\Big)
\end{equation}
and
\begin{eqnarray}\label{otimes.simulation.eq1}
\widehat {\bf X}_{d, M, \otimes}
& \hskip-0.08in = & \hskip-0.08in\frac{1}{2}  \sum_{(i,j)\in {\mathcal S}_M} ({\bf u}_{2, j}^T\widetilde{\bf X}_d {\bf u}_{1, i})
 {\bf u}_{2, j} {\bf u}_{1, i}^T  \nonumber \\
& \hskip-0.08in  & \hskip-0.08in  \quad + ({\bf v}_{2, j}^T\widetilde {\bf X}_d {\bf v}_{1, i})
 {\bf v}_{2, j} {\bf v}_{1, i}^T,
  \end{eqnarray}
 where ${\mathcal S}_M$ contains all pairs $(i,j)$ with $\sigma_{1, i}+\sigma_{2, j}$ being some $\mu_k, 0\le k\le M-1$, one of the first $M$-frequencies in the frequency domain of the GFT ${\mathcal F}_\otimes$.
 Shown in Figure \ref{FourierGFT.fig} are the GFTs
of the  temperature data set ${\bf X}_1$ on January 1st, 2014 and its
bandlimiting approximations $\widehat {\bf X}_{1, M, \square}$ and
 $\widehat {\bf X}_{1, M, \otimes}$ of the noisy temperature data set  $\widetilde {\bf X}_1$  in the frequency domain of the GFTs ${\mathcal F}_\square$ and ${\mathcal F}_\otimes$, where $M=32, c=4$,
 $\|{\bf X}_1\|_F=286.6332$,
 $\|\widehat {\bf X}_{1, M, \square}-{\bf X}_1\|_F=24.1248$ and
 $\|\widehat {\bf X}_{1, M, \otimes}-{\bf X}_1\|_F=23.7247$.
 This shows that the hourly temperature data set ${\bf X}_1$ has about 91.583\% and 91.723\% energy concentrated on
 the first 32 out of total 768 (about 4.167\%) frequencies of the GFTs ${\mathcal F}_\square$ and ${\mathcal F}_\otimes$ respectively.

Define
 the input signal-to-noise ratio
(ISNR) and
the bandlimiting  signal-to-noise ratio (SNR)  by
	$$
{\rm ISNR}(c)= -20 \log_{10} \frac{\|\widetilde {\bf X}-{\bf X}\|_F}{\|{\bf X}\|_F}$$
and
$$
{\rm SNR}(c, M)=-20 \log_{10} \frac{\|\widehat {\bf X}-{\bf X}\|_F}{\|{\bf X}\|_F}, $$
 where  ${\bf X}$ is the original temperature data ${\bf X}_d, 1\le d\le 31$,
 $\widetilde {\bf X}$ is the  noisy temperature data in \eqref{noiseweather.eq},
  $\widehat {\bf X}$ is the bandlimited temperature data in
  \eqref{square.simulation.eq1} or \eqref{otimes.simulation.eq1}.
  Denote the SNR  obtained by \eqref{square.simulation.eq1}  and  \eqref{otimes.simulation.eq1} by  SNR.IV2 and SNR.IV3  respectively.
   Shown in  Tables \ref{ProductDenoisingT1.table} and \ref{ProductDenoisingT1.table2} are the denoising performances of
 the proposed GFTs for different noise levels $c$ and bandlimiting frequency bandwidths $M$, where  the  ISNR,  SNR.IV2 and SNR.IV3 are taken over the average of 100 trials per day and over 31 days.
From Table \ref{ProductDenoisingT1.table}, we observe  that the proposed GFTs ${\mathcal F}_\square$ and ${\mathcal F}_\otimes$ have  similar good performance on denoising the noisy temperature data sets
 collected in the region of Brest,
 and from Table  \ref{ProductDenoisingT1.table2} that the SNR has slow change for larger frequency bandwidth $M\ge 24$ (about $3.125\%$ of the total numbers of frequencies).
 The possible reasons for the second observation  could be that the  temperature data set
  in the region of Brest (France)
  has strong correlation for different hours and  locations, and
   energy of  the original data set is mainly concentrated on the low frequencies of the proposed GFTs, see  Figure \ref{denoise_M32.fig}.
   This demonstrates that the proposed GFTs ${\mathcal F}_\square$ and ${\mathcal F}_\otimes$ could be used to decompose graph signals on product graphs
    into different frequency components and represent those signals with strong correlation efficiently in the frequency domain, cf. Theorems
    \ref{gft1approximation.thm} and \ref{gft2approximation.thm} and see also Figure \ref{FourierGFT.fig}.

\begin{table}[ht]
		\centering
		\caption{ The average ISNR and bandlimiting
SNR 
for  fixed frequency bandwidth $M=32$ and different  noise levels $c$. }
		\begin{tabular} {|c|c|c|c|}
		\hline\hline
c &ISNR &  SNR.IV2  &   SNR.IV3    \\	\hline
   1 & 23.2701   &17.8334 &  17.9590\\
   \hline
   2 &17.2473 &  17.6570 &  17.7780\\
   \hline
   3 &13.7260 &  17.3822 &  17.4981\\
   \hline
   4& 11.2296  & 17.0294  & 17.1400\\
   \hline
    5 &9.2902 &  16.6332  & 16.7344\\
   \hline
    6 &7.7021 &  16.1838   &16.2836\\
   \hline
    7& 6.3647  & 15.7238   &15.8187\\
   \hline
    8  &5.2108 &  15.2610  & 15.3483\\
   \hline\hline
		\end{tabular}
		\label{ProductDenoisingT1.table}
	\end{table}

%
%

\begin{table}[ht]
		\centering
		\caption{
 The average bandlimiting
SNR for different frequency bandwidths $M$ and fixed   noise levels $c=0$ and $4$,  where the average ISNR is 11.2272 for $c=4$.
}
		\begin{tabular} {|c|c|c|c|c|}
		\hline\hline
\multirow{2}{*}{M}&\multicolumn{2}{c|}{c=0}&\multicolumn{2}{c|}{c=4}\\
\cline{2-5}
\multirow{2}{*}{} &  SNR.IV2 &   SNR.IV3 & SNR.IV2 &  SNR.IV3\\
\hline
16 &  16.3048 &  16.6723 &  16.0119  & 16.3485\\
\hline
   24&   17.4866 &  17.4409  & 16.8893 &  16.8667\\
   \hline
   32&   17.8944 &  18.0213 &  17.0320  & 17.1419\\
   \hline
   48&   18.2314  & 18.2260 &  16.8823  & 16.8886\\
   \hline
   64&  18.6767  & 19.3312 &  16.8299  & 17.2950\\
   \hline
  128&   20.5466 &  20.5481 &  16.4607 &  16.4767\\
  \hline
  256&  23.0639 &  23.6230  & 15.0905  & 15.2524\\
 \hline\hline
		\end{tabular}
		\label{ProductDenoisingT1.table2}
	\end{table}


 We also do the simulations to implement the denoising procedures \eqref{square.simulation.eq1} and  \eqref{otimes.simulation.eq1} on the hourly temperature data set  collected at 218 locations
in the United States on
August 1st,  2010 \cite{ncjs22, Yang2022}. Similar to the temperature data set in France,
the data set can be modeled as  signals on the Cartesian product graph  of order $24\times 218$ (about $6.8125$ times the order
$24\times 32$ of the  Cartesian product graph to model temperature data set in France). Our experiments show that
the  time spent on finding the left/right frequency components
of
the GFTs ${\mathcal F}_\square$  and   ${\mathcal F}_\otimes$ are $24.3662$ and $0.294226$ seconds respectively, which are about $283.00$  and $15.57$ times
more than the time spent on finding frequency components when dealing with the temperature data set in France. This reaffirms numerically that
the GFT ${\mathcal F}_\otimes$ has  much lower computational complexity than the GFT ${\mathcal F}_\square$ does for the directed product graph of a large order.
For different noise levels $c$ and frequency bandwidths $M$, our simulations indicate that the proposed GFTs have similar performance on denoising  the U.S. temperature data set
to the one on denoising the temperature data set in the region of Brest (France).

\begin{appendices}

\section{Proof of Theorem \ref{gft1approximation.thm}}
\label{proof2.appendix}

By \eqref{svd.def}, we have
\begin{eqnarray} \label {gft1approximation.thm.pfeq1}
\| {\bf L}_\square {\bf x}\|_2^2 & \hskip-0.08in   = &
\hskip-0.08in  
{\bf x}^T {\bf V} {\pmb \Sigma}^2 {\bf V}^T {\bf x}=\sum_{k=0}^{N-1} \sigma_k^2 ({\bf v}_k^T{\bf x})^2\nonumber \\
& \hskip-0.08in \ge &\hskip-0.08in  \sigma_{M-1}^2\sum_{k=M}^{N-1}  ({\bf v}_k^T{\bf x})^2
\end{eqnarray}
and
\begin{equation} \label {gft1approximation.thm.pfeq2}
\| {\bf L}_\square^T {\bf x}\|_2^2= 
{\bf x}^T {\bf U} {\pmb \Sigma}^2 {\bf U}^T {\bf x}\ge \sigma_{M-1}^2 \sum_{k=M}^{N-1}  ({\bf u}_k^T{\bf x})^2.
\end{equation}
From \eqref{inversegft.eqn1} and \eqref{gft1approximation.thm.eq1},
it follows that
\begin{eqnarray*} 
\hskip-0.12in&\hskip-0.05in&\hskip-0.05in   \|{\bf x}-{\bf x}_{M,\square}\|_2=\frac{1}{2}\left\|\sum_{k=M}^{N-1}({\bf u}_k{\bf u}_k^T+ {\bf v}_k{\bf v}_k^T){\bf x}\right\|_2\nonumber \\
\hskip-0.12in&\le \hskip-0.05in&\hskip-0.05in \frac{1}{2} \left(\sum_{k=M}^{N-1}({\bf u}_k^T{\bf x})^2\right)^{1/2}+ \frac{1}{2}
\left(\sum_{k=M}^{N-1}({\bf v}_k^T{\bf x})^2\right)^{1/2}.\quad
\end{eqnarray*}
This together with  \eqref{lap.def}, \eqref{gft1approximation.thm.pfeq1} and  \eqref{gft1approximation.thm.pfeq2}   completes the proof.

\section{Proof of Theorem \ref{gft2approximation.thm}}
\label{proof3.appendix}

By \eqref{Vsvd.def}, we have
\begin{equation*}  
\| ({\bf L}_1\otimes {\bf I}_{N_2}) {\bf x}\|_2^2  =  
\sum_{i=0}^{N_1-1} \sum_{j=0}^{N_2-1} \sigma_{1, i}^2 \big(({\bf v}_{1, i}\otimes {\bf v}_{2, j})^T{\bf x}\big)^2
\end{equation*}
and
\begin{equation*}  
\| ({\bf I}_{N_1}\otimes {\bf L}_2) {\bf x}\|_2^2  =  
\sum_{i=0}^{N_1-1} \sum_{j=0}^{N_2-1} \sigma_{2, j}^2 \big(({\bf v}_{1, i}\otimes {\bf v}_{2, j})^T{\bf x}\big)^2.
\end{equation*}
This implies that
\begin{eqnarray} \label{gft2approximation.thm.pfeq1}
\hskip-0.28in  & \hskip-0.08in & \hskip-0.08in (\| ({\bf L}_1\otimes {\bf I}_{N_2}) {\bf x}\|_2 + \| ({\bf I}_{N_1}\otimes {\bf L}_2) {\bf x}\|_2)^2\nonumber\\
\hskip-0.28in  &\hskip-0.08in \ge & \hskip-0.08in  \sum_{i=0}^{N_1-1} \sum_{j=0}^{N_2-1} (\sigma_{1, i}+\sigma_{2, j})^2 \big(({\bf v}_{1, i}\otimes {\bf v}_{2, j})^T{\bf x}\big)^2\nonumber\\
\hskip-0.28in  &\hskip-0.08in \ge & \hskip-0.08in  \mu_{M-1}^2 \sum_{(i,j)\notin {\mathcal S}_M}  \big(({\bf v}_{1, i}\otimes {\bf v}_{2, j})^T{\bf x}\big)^2.
\end{eqnarray}
Similarly, we obtain from \eqref{Vsvd.def} that
\begin{eqnarray} \label{gft2approximation.thm.pfeq2}
\hskip-0.28in  & \hskip-0.08in & \hskip-0.08in (\| ({\bf L}_1^T\otimes {\bf I}_{N_2}) {\bf x}\|_2 + \| ({\bf I}_{N_1}\otimes {\bf L}_2^T) {\bf x}\|_2)^2\nonumber\\
\hskip-0.28in  &\hskip-0.08in \ge & \hskip-0.08in  \mu_{M-1}^2 \sum_{(i,j)\notin {\mathcal S}_M}  \big(({\bf u}_{1, i}\otimes {\bf u}_{2, j})^T{\bf x}\big)^2.
\end{eqnarray}
From \eqref{inverse2gft.eq} and \eqref{gft2approximation.thm.eq1}
it follows that
\begin{eqnarray} \label {gft2approximation.thm.pfeq3}
\hskip-0.05in&\hskip-0.05in&\hskip-0.05in   \|{\bf x}-{\bf x}_{M,\otimes}\|_2\le
\Big(\sum_{(i,j)\notin {\mathcal S}_M}  \big(({\bf u}_{1, i}\otimes {\bf u}_{2, j})^T{\bf x}\big)^2\Big)^{1/2}\nonumber\\
\hskip-0.05in&\hskip-0.05in&\hskip-0.05in   \quad +
\Big(\sum_{(i,j)\notin {\mathcal S}_M}  \big(({\bf v}_{1, i}\otimes {\bf v}_{2, j})^T{\bf x}\big)^2\Big)^{1/2}.
\end{eqnarray}
Combining \eqref{gft2approximation.thm.pfeq1}, \eqref{gft2approximation.thm.pfeq2}  and  \eqref{gft2approximation.thm.pfeq3} establishes the  desired estimate in \eqref{gft2approximation.thm.eq2}.

\section {Proof of Theorem \ref{sameGFT.thm}}
\label{proof.appendix}

For $l=1, 2$, let ${\bf U}_l$ and ${\bf V}_l$ be  orthogonal matrices in the SVD \eqref{Vsvd.def} of Laplacians ${\bf L}_l$ on the graphs ${\mathcal G}_l$,
and write $
{\bf U}_l=[u_l(i,j)]_{0\le i, j\le N_l-1}, {\bf V}_l=[v_l(i,j)]_{0\le i, j\le N_l-1}$ and ${\bf L}_l=[a_l(i,j)]_{0\le i, j\le N_l-1}$.
By the assumption on GFTs ${\mathcal F}_\square$ and ${\mathcal F}_\otimes$,
the Laplacian ${\bf L}_\square$ in \eqref{lap.def} has the following  decomposition
\begin{equation}\label{proof.eq1}
{\bf L}_\square = {\bf U}_\otimes  {\pmb \Sigma} {\bf V}_\otimes^T,
\end{equation}
where  ${\bf U}_\otimes$ and ${\bf V}_\otimes$ are given in \eqref{orthogonalMatrix2.def} and
$\pmb \Sigma$ is a diagonal matrix with nonnegative diagonal entries which are not necessarily in a nondecreasing order.
Let $\delta(i,j), 0\le i, j\le N_1-1$, be the Kronecker delta and write ${\pmb \Sigma}={\rm diag}({\pmb \Sigma}_0, \ldots, {\pmb \Sigma}_{N_1-1})$, where ${\pmb \Sigma}_i, 0\le i\le N_1-1$,
are diagonal matrices of size $N_2$.
By \eqref{lap.def} and \eqref{proof.eq1}, we have
\begin{eqnarray}\label{proof.eq2}
\hskip-0.25in & \hskip-0.08in & \hskip-0.08in \sum_{k=0}^{N_1-1} u_1(i,k) v_1(j,k) {\bf U}_2 {\pmb \Sigma}_k {\bf V}_2^T\nonumber\\
\hskip-0.25in& \hskip-0.08in = & \hskip-0.08in  a_1(i,j) {\bf I}_{N_2}+ \delta(i,j) {\bf L}_2, \  0\le i, j\le N_1-1.
\end{eqnarray}

 Let ${\mathbb R}^{N_2\times N_2}$ be the Hilbert space of the real matrices of size $N_2\times N_2$ with the inner product
 of two matrices  $ {\bf A}=[a(i,j)]_{0\le i, j\le N_2-1}$ and ${\bf B}=[b(i,j)]_{0\le i, j\le N_2-1}$
defined by
\begin{equation*}\langle{\bf  A}, {\bf B}\rangle=\sum_{i,j=0}^{N_2-1} a(i,j)b(i,j).\end{equation*}
Write
\begin{equation} \label{proof.eq3}
{\bf U}_2 {\pmb \Sigma}_k {\bf V}_2^T= b_k {\bf I}_{N_2}+c_k {\bf L}_2+ {\bf W}_k, 0\le k\le N_1-1,\end{equation}
where $b_k, c_k\in {\mathbb R}$ and ${\bf W}_k, 0\le k\le N_1-1$, are orthogonal to the linear subspace of ${\mathbb R}^{N_2\times N_2}$ spanned by
${\bf I}_{N_2}$ and ${\bf L}_2$.
By \eqref{proof.eq2} and \eqref{proof.eq3}, we have
$$ \sum_{k=0}^{N_1-1} u_1(i,k) v_1(j,k) {\bf W}_k={\bf O}_{N_2}, \ 0\le i, j\le N_1-1.
$$
This together with the orthogonal properties of matrices ${\bf U}_1$ and ${\bf V}_1$ implies that
\begin{equation}
\label{proof.eq4}
{\bf W}_k={\bf O}_{N_2},\  0\le k\le N_1-1.
\end{equation}

By the non-edgeless assumption on the graph ${\mathcal G}_2$,
the unit matrix
${\bf I}_{N_2}$ and the Laplacian ${\bf L}_2$ on the graph ${\mathcal G}_2$ are linearly independent in ${\mathbb R}^{N_2\times N_2}$.  Therefore combining \eqref{proof.eq2}, \eqref{proof.eq3} and \eqref{proof.eq4}, we obtain
\begin{equation}\label{proof.eq5}
\sum_{k=0}^{N_1-1} u_1(i,k) v_1(j,k) b_k =a_1(i,j)
\end{equation}
and
\begin{equation}\label{proof.eq6}
\sum_{k=0}^{N_1-1} u_1(i,k) v_1(j,k) c_k
=  \delta(i,j),
\end{equation}
where $0\le i, j\le N_1-1$.
Let ${\bf B}$ and ${\bf C}$ be the diagonal matrices with diagonal entries $b_k$ and $c_k, 0\le k\le  N_1-1$. Then we can rewrite
\eqref{proof.eq5} and \eqref{proof.eq6} in the following matrix formulation:
\begin{equation} \label{proof.eq7}
{\bf U}_1{\bf B} {\bf V}_1^T={\bf L}_1\ \ {\rm and}\ \ {\bf U}_1 {\bf C} {\bf V}_1^T={\bf I}_{N_1}.
\end{equation}
This implies that
${\bf L}_1= {\bf U}_1 {\bf B} {\bf C}^{-1} {\bf U}_1^T$
is symmetric. Hence the graph ${\mathcal G}_1$ is undirected.

By the non-edgeless assumption on the graphs ${\mathcal G}_1$,
the unit matrix
${\bf I}_{N_1}$ and the Laplacian ${\bf L}_1$ on the graph ${\mathcal G}_1$ are linearly independent.
Then  we conclude from \eqref{proof.eq7} that the diagonal matrix ${\bf B}{\bf C}^{-1}$
is not a multiple of the identity matrix ${\bf I}_{N_1}$, which in turn implies that
there exist $0\le k_1\ne k_2\le N_1-1$ such that $(b_{k_1}, c_{k_1})$ and
$(b_{k_2}, c_{k_2})$ are  linearly independent in ${\mathbb R}^2$. Hence there are two diagonal matrices
$\tilde {\bf B}$ and $\tilde {\bf C}$ by \eqref{proof.eq2}, \eqref{proof.eq3} and \eqref{proof.eq4} such that
$$
{\bf U}_2\tilde {\bf B} {\bf V}_2^T={\bf L}_2 \ \ {\rm and} \ \ {\bf U}_2 \tilde {\bf C} {\bf V}_2^T={\bf I}_{N_2}.$$
Therefore
${\bf L}_2= {\bf U}_2 \tilde {\bf B} (\tilde {\bf C})^{-1} {\bf U}_2^T$
is symmetric. This completes the proof that the graph ${\mathcal G}_2$ is undirected.

\end{appendices}

\end{document}